\documentclass[notitlepage,english,aps,floats,onecolumn,showpacs,nofootinbib,floatfix]{revtex4-2}

\usepackage{pslatex}
\usepackage[T1]{fontenc}
\usepackage[latin1]{inputenc}
\usepackage[dvips]{graphicx}
\usepackage{epsfig}
\usepackage{longtable}
\usepackage{float}
\usepackage{calc}
\usepackage{ifthen}
\usepackage{amsmath}
\usepackage{hyperref}
\usepackage{amssymb}

\usepackage{color}

{
{
{
\newcommand{\bea}{\begin{eqnarray}}
\newcommand{\eea}{\end{eqnarray}}

\newcommand{\nc}{\newcommand}
\nc{\renc}{\renewcommand}
\nc{\eqs}[2]{\mbox{Eqs.~(\ref{#1},\,\ref{#2})}}
\nc{\eq}[1]{\mbox{Eq.~(\ref{#1})}}
\nc{\figs}[2]{\mbox{Figs.~(\ref{#1},\,\ref{#2})}}
\nc{\fig}[1]{\mbox{Fig~.(\ref{#1})}}
\nc{\be}[1]{\begin{equation} \mbox{$\label{#1}$}}
\nc{\ee}{\vspace{0.1cm}\end{equation}}

\newcommand{\bean}{\begin{eqnarray*}}
\newcommand{\eean}{\end{eqnarray*}}

\def\GeV{{\rm \ GeV}}

\def\lae{\;^{<}_{\sim} \;} \def\gae{\; ^{>}_{\sim} \;}

\begin{document}

\title{A Minimal Approach to Baryogenesis via Affleck-Dine and Inflaton Mass Terms}

\author{Amy Lloyd-Stubbs  and John McDonald }
\email{a.lloyd-stubbs@lancaster.ac.uk}
\email{j.mcdonald@lancaster.ac.uk}
\affiliation{Dept. of Physics,  
Lancaster University, Lancaster LA1 4YB, UK}

\begin{abstract}

We present a minimal approach to the generation of the baryon ($B$) asymmetry of the Universe, in which the asymmetry is generated in a complex inflaton condensate via $B$-violating quadratic inflaton potential terms and the Affleck-Dine (AD) mechanism. We show that the $B$-violating quadratic mass terms create an oscillating asymmetry in the complex inflaton condensate at late times. The final asymmetry transferred to the Standard Model sector at reheating is naturally reduced to the magnitude of the observed $B$ asymmetry by the effect of averaging over the $B$ oscillations. This approach to baryogenesis can easily be  realised in a wide range of inflation models.

\end{abstract}
 
\maketitle

\section{Introduction}   

The Affleck-Dine (AD) mechanism \cite{ad,ad2} provides a remarkably simple and elegant explanation for the baryon  ($B$) asymmetry of the Universe. A complex scalar with a $U(1)$ global symmetry, corresponding to conserved baryon number, evolves into a coherently oscillating condensate. 
$B$ violating terms in the potential act on the field, pushing it into an elliptical trajectory in the complex field plane, which is equivalent to a $B$ asymmetry in the scalar field.   

The conventional AD mechanism is based on a complex scalar field $\Phi$ with a potential which at late times is dominated by a $|\Phi|^2$ mass term. Higher-order operators that violate baryon number cause the real ($\phi_{1}$) and imaginary ($\phi_{2}$) parts of $\Phi$ to evolve differently when the $|\Phi|^{2}$ term comes to dominate the potential, pushing the trajectory into an ellipse in the $(\phi_{1}, \phi_{2})$ plane. The higher-order operators become less important as the magnitude of $\Phi$ decreases due to expansion, effectively switching off the $B$ violation and leaving a conserved baryon asymmetry in the complex field at late times. 

Here we will present a new and unconventional implementation of AD baryogenesis, in which $B$-violating $\Phi^{2}$ terms in the potential of a complex inflaton $\Phi$ generate the asymmetry\footnote{The same model can also be used to generate a lepton asymmetry which is subsequently processed via sphalerons into a baryon asymmetry.}. (Applications of the conventional AD mechanism to a complex inflaton have been considered in \cite{adc3,ea1,adc1,adc2,adc4}.) We will show that these terms generate a $B$ asymmetry in the $\Phi$ condensate which oscillates about zero. When the condensate asymmetry is transferred to the Standard Model (SM) sector by $\Phi$ decay, a net asymmetry is left in the SM sector. The oscillating baryon asymmetry initially generated in the $\Phi$ condensate is typically much larger than that required to explain the observed baryon-to-entropy ratio. The asymmetry transferred to the SM is subsequently suppressed by averaging over the condensate asymmetry oscillations, reducing the asymmetry to the observed value\footnote{AD baryogenesis via mass terms has previously been considered in the context of a different class of model in \cite{gor}. The analysis of \cite{gor} assumes that the averaging over of asymmetry oscillations washes out the final asymmetry. We will show that although the asymmetry is suppressed, it is significantly non-zero. This suppression plays an important role in the model described here.}. The resulting model is dynamically quite different from existing inflaton-based AD baryogenesis models, with the inflaton asymmetry being generated at late times during inflaton oscillations rather than during or shortly after inflation.  

The paper is organised as follows. In Section 2 we discuss the generation of the asymmetry via quadratic B-violating potential terms. In Section 3 we consider possible washout of the asymmetry via inflaton exchange operators. In Section 4 we discuss the validity of the classical calculation of the asymmetry. In Section 5 we present our conclusions. 

\section{Affleck-Dine Baryogenesis via Quadratic Potential Terms} 

We will consider a renormalisable $B$ symmetric inflaton potential together with $B$-violating $\Phi^2$ terms,
\be{e1}  V(\Phi) = m_{\phi}^{2} |\Phi|^{2}  + \lambda_{\Phi} |\Phi|^{4} - (A \Phi^2 + {\rm h.\,c.})   ~, \ee
where $A$ is real and positive. Such potentials are naturally compatible with inflation models which are non-minimally coupled to gravity \cite{bbs}. More generally, they represent the leading order terms of an inflaton potential during post-inflation evolution\footnote{Whilst the inflaton is the natural candidate for the field responsible for reheating, we note that the model can apply to any coherently oscillating complex scalar that is responsible for reheating.}. $\Phi$ is initially coherently oscillating, with the potential dominated by the $|\Phi|^4$ term and with no asymmetry in the field. In terms of $\Phi = (\phi_{1} + i \phi_{2})/\sqrt{2}$, the potential becomes
\be{e4} V(\Phi) = \frac{1}{2} (m_{\Phi}^{2} - 2A) \phi_{1}^{2} 
+ \frac{1}{2} (m_{\Phi}^{2} + 2A) \phi_{2}^{2} + \frac{\lambda_{\Phi}}{4} (\phi_{1}^{2} + \phi_{2}^{2})^{2}  ~.\ee
The field equations are
\be{e5a} \ddot{\phi}_{1} + 3 H \dot{\phi}_{1} 
  = -  m_{1}^{2} \phi_{1} - \lambda_{\Phi}(\phi_{1}^{2} + \phi_{2}^{2}) \phi_{1} ~\ee
and
\be{e5b} \ddot{\phi}_{2} + 3 H \dot{\phi}_{2}  = -  m_{2}^{2} \phi_{2} - \lambda_{\Phi}(\phi_{1}^{2} + \phi_{2}^{2}) \phi_{2} ~,\ee
where 
\be{e6a} m_{1}^{2} = m_{\Phi}^{2} - 2A  \;\;\;;\;\;\; 
 m_{2}^{2} = m_{\Phi}^{2} + 2A   ~.\ee
In the limit $\lambda_{\Phi} \rightarrow 0$ the equations for $\phi_{1}$ and $\phi_{2}$ are decoupled from each other, with coherently oscillating solutions for $\phi_{1}$ and $\phi_{2}$ which have angular frequencies $m_{1}$ and $m_{2}$, respectively.  

We first derive an analytical expression for the asymmetry using a threshold approximation, which we then compare to a complete numerical solution. In the threshold approximation we consider the potential to be approximated by 
$$ V(\Phi) =  \lambda_{\Phi} |\Phi|^{4}  \;\;\;;\;\;\; \phi > \phi_{*}   $$
\be{ex2}  V(\Phi) = m_{\Phi}^{2} |\Phi|^{2} - (A \Phi^{2} + {\rm h.c.} )    \;\;\;;\;\;\; \phi < \phi_{*}   ~,\ee 
where $\phi_{*} = m_{\Phi}/\sqrt{\lambda_{\Phi}}$ is the value of $\phi$ at which $V'(\phi)$ becomes dominated by the $|\Phi|^4$ term (here $\Phi = \phi e^{i \theta}/\sqrt{2}$ and we have set $A = 0$ when determining $\phi_{*}$). 
The potential is initially strongly dominated by the $|\Phi|^4$ term, with $\phi_{i} \gg \phi_{*}$, and the field is initially at rest with initial values $(\phi_{1,\;i}, \phi_{2,\;i})$. 
Assuming rapid coherent oscillations,  the field amplitude will initially evolve as $\phi \propto 1/a$ when $\phi > \phi_{*}$. Therefore 
the field amplitudes at $a_{*}$ are  
\be{ex5} \phi_{1,\;*} = \left(\frac{a_{i}}{a_{*}}\right)\phi_{1,\;i} = \left(\frac{\phi_{*}}{\phi_{i}}\right) \phi_{1,\;i}  \;\;\; ; \;\;\;  \phi_{2,\;*} = \left(\frac{a_{i}}{a_{*}}\right)\phi_{2,\;i} = \left(\frac{\phi_{*}}{\phi_{i}}\right) \phi_{2,\;i}   ~,\ee
where $ \phi_{i} = \left(\phi_{1,\;i}^{2} + \phi_{2,\;i}^{2}\right)^{1/2} $.
The field evolves purely due to the mass squared terms once $a > a_{*}$. We assume that $m_{1,2} \gg H$, so that we can neglect the effect of expansion on the rapid $\phi_{1,2}$ oscillations and simply factor in the effect of expansion by damping the oscillation amplitude. The solution for $\phi_{1}$ and $\phi_{2}$ is then 
\be{e7} \phi_{1} = \phi_{1,\;*} \left(\frac{a_{*}}{a}\right)^{3/2} \cos(m_{1} (t - t_{*}))  
\;\;\; ; \;\;\;  \phi_{2} = \phi_{2,\;*} \left(\frac{a_{*}}{a}\right)^{3/2} \cos(m_{2} (t - t_{*} ))  ~.\ee    
The baryon asymmetry in the $\Phi$ condensate is  
\be{e9} n(t) = i \left(\Phi^{\dagger}  \dot{\Phi} -  \dot{\Phi}^{\dagger} \Phi \right)   = \dot{\phi}_{1} \phi_{2} - \dot{\phi}_{2} \phi_{1}  ~.\ee
Therefore 
\be{e10} n(t) = \phi_{1,\;*} \phi_{2,\;*}  \left(\frac{a_{*}}{a}\right)^{3} \left[m_{2} \sin(m_{2}(t - t_{*})) \cos(m_{1} (t - t_{*}) ) - m_{1} \sin(m_{1} (t - t_{*}) ) \cos(m_{2} (t - t_{*})) \right]   ~.\ee 
We will assume that $2 A \ll m_{\Phi}^{2}$. In this limit, to leading order in $A/m_{\Phi}^{2}$,  the 
condensate baryon asymmetry becomes
\be{e11} n(t) = \phi_{1,\;*} \phi_{2,\;*}  \left(\frac{a_{*}}{a}\right)^{3} \left[ m_{\Phi} \sin\left( \frac{2 A (t - t_{*}) }{m_{\Phi}} \right) + \frac{A}{m_{\Phi}} \sin \left(2 m_{\Phi} (t - t_{*})  \right) \right]   ~.\ee
During averaging over the $\phi_{1\,,2}$ coherent field oscillations, we can consider the 
scale factor to be constant since $H \ll m_{\Phi}$. The second term in \eq{e11} then averages to zero. The condensate asymmetry at $t > t_{*}$, in terms of the initial field values, is then  
\be{e13}  n(t) = \phi_{1,\;i} \phi_{2,\;i} \left(\frac{\phi_{i}}{\phi_{*}}\right) \left(\frac{a_{i}}{a}\right)^{3} m_{\Phi} \sin\left( \frac{2 A (t - t_{*}) }{m_{\Phi}} \right)  ~. \ee 
Thus the baryon asymmetry in the $\Phi$ condensate oscillates about zero with period $T_{asy} = \pi m_{\Phi}/A$. 

It is useful to define a comoving asymmetry $n_{c}(t) \equiv (a(t)/a_{i})^{3} n(t)$, which is constant when there is no production or decay of the asymmetry. For the threshold model at $t > t_{*}$
\be{e15} n_{c}(t) = \phi_{1,\;i} \phi_{2,\;i} \left(\frac{\phi_{i}}{\phi_{*}}\right) m_{\Phi} \sin\left( \frac{2 A (t - t_{*}) }{m_{\Phi}} \right)  ~,\ee
with $n_{c}(t) = 0$ at $t < t_{*}$.
The $\Phi$ condensate asymmetry is assumed to transfer to a conserved SM baryon asymmetry via $B$-conserving $\Phi$ decays to SM particles\footnote{A specific implementation of the model to baryogenesis from AD leptogenesis via a decaying inflaton will be presented in a future work \cite{aj2}. Here we focus on the general features of inflaton mass term AD baryogenesis.}\textsuperscript{,}\footnote{It is also possible for the inflaton to decay via gravity mediated modes \cite{grav}. The importance of this process will depend upon the coupling of the inflaton to the Ricci curvature in a given inflation model.} The condensate will decay away completely after a time $ t_{R} \approx \Gamma_{\Phi}^{-1}$, where $R$ denotes reheating, with continuous production of SM baryon asymmetry due to decay of the condensate asymmetry from $t_{*}$ to $t_{R}$. Neglecting any reduction of the $\Phi$ field due to decays at $t < t_{R}$, the comoving baryon asymmetry transferred to the SM sector, which we denote by $\hat{n}_{c}$(t),  is  
\be{e17}  \hat{n}_{c}(t) = \int_{t_{i}}^{t} \Gamma_{\Phi} n_{c}(t) dt   ~.\ee 
Thus the comoving baryon asymmetry transferred out of the $\Phi$ condensate as a function of $t$ is 
\be{e20} \hat{n}_{c}(t) = \frac{\Gamma_{\Phi} \phi_{1,\;i}\phi_{2,\;i}  m_{\Phi}^{2}}{2 A} \left(\frac{\phi_{i}}{\phi_{*}}\right)\left[1 - \cos\left(\frac{2 A (t-t_{*}) }{m_{\Phi}} \right) \right]   ~.\ee 
$\hat{n}_{c}(t)$ increases linearly with $t - t_{*}$ until $t - t_{*} \approx \pi m_{\Phi}/ 4 A$. On longer timescales,  $\hat{n}_{c}(t)$  oscillates between a maximum value and zero with period $T_{asy}$. The maximum possible asymmetry is obtained when $A = A_{max} = \pi m_{\Phi} \Gamma_{\Phi}/2$.

The $\Phi$ condensate decays away completely once $t - t_{*} \gae  \Gamma_{\Phi}^{-1}$. To take into account the B-conserving decay of the condensate asymmetry, we include in \eq{e17} an exponential decay factor, 
\be{e20a} \hat{n}_{c}(t) = \int_{t_{*}}^{t} \Gamma_{\Phi} n_{c}(t) e^{-\Gamma_{\Phi}(t - t_{*})}  dt   ~.\ee 
The total comoving asymmetry transferred to the SM sector as $t \rightarrow \infty$ 
is then
\be{e21}  \hat{n}_{c,\;tot} = \frac{\Gamma_{\Phi} \phi_{1,\;i}\phi_{2,\;i} m_{\Phi}^{2}}{2 A} \left(\frac{\phi_{i}}{\phi_{*}}\right)\left(1 + \left(\frac{\Gamma_{\Phi} m_{\Phi}}{2 A}\right)^{2} \right)^{-1}  ~.\ee 
The transferred asymmetry is proportional to $A$ until $A > \Gamma_{\Phi} m_{\Phi}/2$, in which case $\tau_{\Phi} > T_{asy}$ and the transferred asymmetry decreases as $A^{-1}$ and $\tau_{\Phi}^{-1}$, where $\tau_{\Phi} = \Gamma_{\Phi}^{-1}$ is the lifetime of the $\Phi$ scalars. This can be understood as due to the effect of averaging condensate oscillations over the time taken for the condensate to decay. When $\tau_{\phi} \gg T_{asy}$, the asymmetry in the condensate will undergo many oscillations from positive to negative values during the decay of the condensate. Therefore the asymmetry produced during a positive half-cycle will almost cancel against that produced during the following negative 
half-cycle, up to the effect of the small decrease in the condensate asymmetry amplitude due to the decay of the condensate during $\Delta t \sim T_{asy}$. Therefore only a small net asymmetry is produced during each condensate oscillation cycle as compared to the case with $T_{asy} \gae \tau_{\Phi}$, where there is no averaging over oscillations.  

  We first consider the case where the lifetime of $\Phi$ is much longer than $T_{asy}$, such that $2 A/m_{\Phi} \Gamma_{\Phi} \gg 1$. $\hat{n}_{c\;tot}$ can then be expressed as 
\be{e25} \hat{n}_{c,\;tot} = \frac{ \Gamma_{\Phi} \phi_{i}^{2} m_{\Phi}^{2} \sin \left(2 \theta\right) }{4 A}   \left(\frac{\phi_{i}}{\phi_{*}}\right)  ~, \ee 
where $\theta$ is the initial phase of $\Phi$. 
The total baryon asymmetry transferred to the SM, $\hat{n}_{tot}$,  is then
\be{e26a}  \hat{n}_{tot} = \left( \frac{a_{i}}{a_{R}}\right)^{3} \hat{n}_{c,\;tot}  = \frac{3 M_{Pl}^{2} \Gamma_{\Phi}^{3} \sin \left(2 \theta\right)   }{2 A} ~,\ee
where we have used $a \propto H^{-2/3}$ when $a > a_{R}$ and $a \propto 1/\phi$ when $a < a_{R}$ to obtain the final expression. 
This can also be expressed in terms of the baryon-to-entropy ratio, $n_{B}/s$. Using $s = 4  k_{T}^{2} T^{3}$ and $\Gamma_{\Phi} = H_{R} = k_{T_{R}} T_{R}^{2}/M_{Pl}$, where $T_{R}$ is the reheating temperature and $k_{T} = (\pi^{2} g(T)/90)^{1/2}$, the baryon-to-entropy ratio is 
\be{e26b} \frac{n_{B}}{s} \equiv \frac{\hat{n}_{tot}}{s} = \frac{3}{8} \frac{k_{T_{R}} T_{R}^{3}\sin \left(2 \theta\right)  }{A M_{Pl} }    = 5.2 \times 10^{-21}   \frac{m_{\Phi}^{2}}{A}    \left( \frac{T_{R}}{10^{8} \GeV} \right)^{3}  \left( \frac{10^{13} \GeV}{m_{\Phi}} \right)^{2}  \sin \left(2 \theta\right)                                       ~,\ee
where we have normalised the expression to some representative values\footnote{$T_{R} = 10^{8} \GeV$ is within the range of reheating temperatures that may be detectable in the spectrum of primordial gravitational waves \cite{pgw}.} of $T_{R}$ and $m_{\Phi}$. 
The observed baryon-to-entropy ratio is  
$ (n_{B}/s)_{obs}  = 0.861 \pm 0.005 \times 10^{-10} $.  
In order to account for the observed asymmetry, we require that
\be{e40} \frac{A^{1/2}}{m_{\Phi}} = 7.8 \times 10^{-6} \sin^{1/2} \left(2 \theta\right)    
\left( \frac{10^{13} \GeV }{m_{\Phi}} \right) 
\left( \frac{T_{R}}{10^{8} \GeV} \right)^{3/2} 
~.\ee
The  
 maximum possible asymmetry, which corresponds to $A = \Gamma_{\Phi} m_{\Phi}/2$ in \eq{e21}, is  
\be{ea1} \frac{n_{B,\; max}}{s} = \frac{3 T_{R} \sin \left(2 \theta\right)  }{8 m_{\Phi}} = 3.8 \times 10^{-6} \, \left( \frac{T_{R}}{10^{8} \GeV} \right)  \left( \frac{10^{13} \GeV}{m_{\Phi}} \right) \sin \left(2 \theta\right)    ~.\ee
This can easily be much larger than the observed baryon asymmetry. Therefore the suppression of the asymmetry by   
averaging over oscillations plays an important role in this model. 

In the case where $\Phi$ decays before any condensate asymmetry oscillations can occur, corresponding to $\Gamma_{\Phi} m_{\Phi}/2A \gg 1$ in \eq{e21}, the total transferred asymmetry obtains an additional factor $(2 A /\Gamma_{\Phi} m_{\Phi})^{2}$ compared to \eq{e26b}. Therefore  
\be{ea1a} \frac{n_{B}}{s} = \frac{3}{2} \frac{A M_{Pl}\sin(2 \theta)}{k_{T_{R}} T_{R} m_{\phi}^{2}}  ~\ee 
and we find that the required value of
 $A^{1/2}/m_{\Phi}$ is   
\be{ea2} \frac{A^{1/2}}{m_{\Phi}} = 8.9 \times 10^{-11} \,\left( \frac{T_{R}}{10^{8} \GeV} \right)^{1/2}
\left(\frac{1}{\sin \left(2 \theta\right)} \right)^{1/2}   
    ~.\ee  
This is typically much smaller than in the case with asymmetry oscillations, due to the lack of additional suppression of the baryon asymmetry from averaging over condensate oscillations.

The threshold asymmetry is a good approximation if the $B$-violating mass terms do not cause the field to significantly evolve until the potential is $|\Phi|^{2}$ dominated. The condition for this to be true, which we have confirmed in our numerical solutions,  is that the mass of the angular field perturbations about the minimum of the potential as a function of $\theta$, $m_{\delta \theta} = 2 A^{1/2}$, is less than $H$ when $\phi = \phi_{*}$. This is satisfied if 
\be{e41}  \frac{A^{1/2}}{m_{\Phi}} \lae \frac{A^{1/2}_{th}}{m_{\Phi}} = \frac{m_{\Phi}}{4 \sqrt{\lambda_{\Phi}} M_{Pl}} \equiv 1.0 \times 10^{-6} \lambda_{\Phi}^{-1/2} \left( \frac{m_{\Phi}}{10^{13} \; \GeV} \right)  ~.\ee

 We finally compare the threshold approximation to the complete numerical solution for the case $\Gamma_{\Phi} (t - t_{*}) \ll 1$ \footnote{Further details of the numerical analysis will be presented in \cite{aj2}}. As an example, we show in Figure 1 the numerical results for the case $m_{\Phi} = 10^{16} \GeV$ and $\lambda_{\Phi} = 0.1$ for a range of values of $A^{1/2}/m_{\Phi}$. The analytical approximation in left-hand figure is given by \eq{e15} and in the right-handed figure by \eq{e20}, with $\Gamma_{\Phi}$ corresponding to $T_{R} = 10^{8}\GeV $. For this case, the upper limit for the threshold approximation to be valid  
is $A_{th}^{1/2}/m_{\Phi} \approx 3 \times 10^{-3}$.  We find that the threshold approximation is in perfect agreement with the numerical solution for both the condensate and transferred asymmetries when \eq{e41} is satisfied. For larger $A^{1/2}/m_{\Phi}$, the evolution during the $|\Phi|^4$ dominated era modifies the asymmetries. The amplitude of the transferred asymmetry $A \hat{n}_{c}$ rapidly decreases with increasing $A > A_{th}$ down to an approximately constant value,  which is suppressed relative to the threshold value of $A \hat{n}_{c}$ by a factor that numerically is approximately $m_{\Phi}/10^{17} {\rm GeV}$. The transferred asymmetry $A \hat{n}_{c}$ oscillates between zero and a maximum when the threshold approximation is valid,  but for larger $A^{1/2}/m_{\Phi}$ it oscillates about zero. However, since the transferred asymmetry is the {\it total}  asymmetry transferred to the SM sector as a function of time after averaging over condensate asymmetry oscillations, the oscillation of the transferred asymmetry about zero has no impact on the typical magnitude of the baryon asymmetry transferred to the SM.

\vspace{-2.5cm}

\begin{figure}[H]
\begin{center}
\hspace*{-1.0cm}\includegraphics[trim = -3.5cm 0cm 0cm 0cm, clip = true, width=0.55\textwidth, angle = -90]{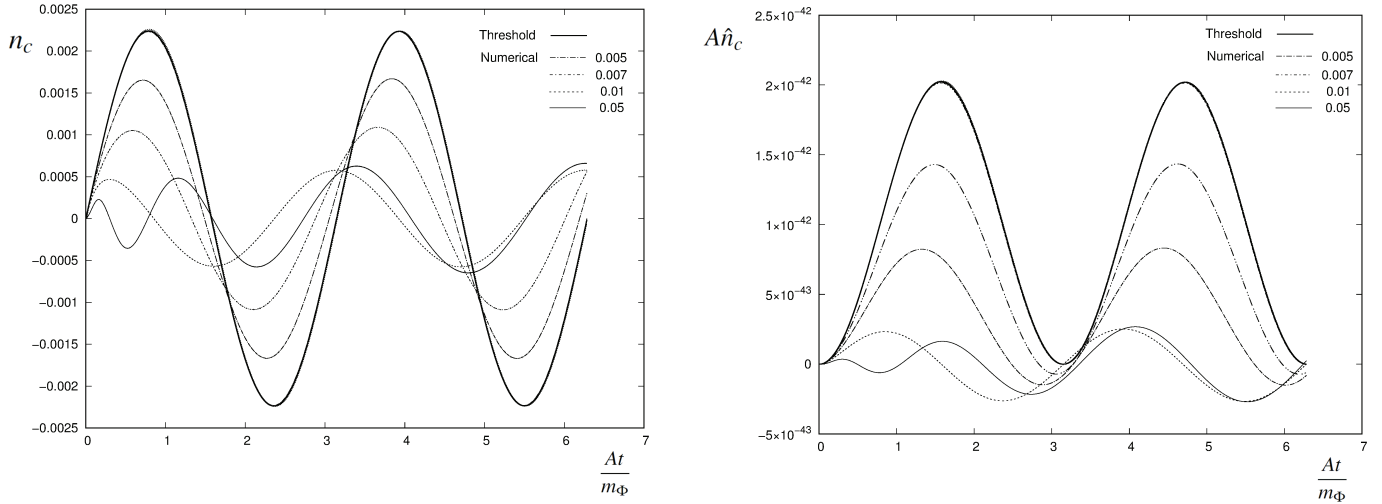}
\caption{The condensate asymmetry (left) and transferred asymmetry (right) for the case $m_{\Phi} = 10^{16} \GeV$, $\lambda_{\Phi} = 0.1$ and $T_{R} = 10^{8}\GeV $. The threshold asymmetry and the numerical results for $A^{1/2}/m_{\Phi} = 0.001, 0.005, 0.007, 0.01$ and 0.05 are shown. (The numerical result for $A^{1/2}/m_{\Phi} = 0.001$ coincides with the threshold result, in agreement with \eq{e41}.) } 
\label{fig1}
\end{center}
\end{figure}

\section{Baryon Washout due to Inflaton Exchange} 

In application to a specific model, the possible washout of the asymmetry must be considered. The interaction which allows the decay of the inflaton will generally result in a B-violating operator via $\Phi$ exchange. Dimensionally, the rate of B-violating scattering processes at reheating due to $\phi_{1}$ and $\phi_{2}$ exchange is
\be{r1}  \Gamma_{\Delta B} \sim \frac{\lambda_{\psi}^{2} A^{2} T_{R}^{5}}{m_{\Phi}^{8}}    ~,\ee
where $\lambda_{\psi}$ is the coupling responsible for $\Phi$ decay and $A$ is necessary in the scattering amplitude in order to have B-violation. Washout due to $\Phi$ exchange will be negligible if  $\Gamma_{\Delta B} < H(T_{R})$, which is satisfied if 
\be{r2}    \lambda_{\psi} \lesssim \frac{m_{\Phi}^{4}}{M_{Pl}^{1/2} T_{R}^{3/2} A} =   6 \times 10^{4} \left(\frac{m_{\Phi}^{2}}{A}\right) \left(\frac{m_{\Phi}}{10^{13} \GeV}\right)^{2} \left(\frac{10^{8} \GeV}{T_{R}}\right)^{3/2}    ~. \ee
The inflaton decay rate is $\Gamma_{\Phi} \approx \lambda_{\psi}^{2} m_{\Phi}/4 \pi$, therefore the reheating temperature from $H(T_{R}) = \Gamma_{\Phi}$ is 
$T_{R} \approx \lambda_{\psi} (m_{\Phi} M_{Pl})^{1/2}$. Thus \eq{r2} is satisfied if 
\be{r3} T_{R} \lae \left( \frac{m_{\Phi}^{2}}{A} \right)^{2/5} m_{\Phi}  ~, \ee 
where $A < m_{\Phi}^{2}$. Therefore washout due to B-violating $\Phi$ exchange is negligible if $T_{R} \lae m_{\Phi}$ and so it is unlikely present a serious obstacle to this class of model. A complete analysis of washout will depend upon the specific model for the decay of the inflaton and the transfer of the baryon asymmetry.

\section{Validity of the Classical Analysis of the Baryon Asymmetry}

Throughout our analysis we have assumed that  classical fields can be used to calculate the baryon asymmetry. When the potential is dominated by quadratic terms, the $\phi_{1}$ and $\phi_{2}$ fields evolve as independent non-interacting coherently oscillating scalars. In general, a classical oscillating scalar field corresponds to a quantum coherent state in the limit where the occupation number of the state is large compared to one \cite{david,loz}. The condition for this to be true is that $\phi_{i} > m_{\Phi}$ ($i = 1,\,2$). 
However, this is typically not satisfied at inflaton decay in the present model. Nevertheless, the classical calculation of the baryon asymmetry remains correct. This is because it is the coherent state corresponding to the classical field that is important for AD baryogenesis.

By construction, the expectation value of the field operator $\hat{\phi}_{i}$ in the coherent state $|\phi_{i}(t)>$ is equal to the classical field $\phi_{i,\;cl}(t)$
\be{cs1}   <\phi_{i}(t)| \hat{\phi}_{i} |\phi_{i}(t)> = \phi_{i,\;cl}(t)     ~.\ee 
We have included a time dependence in the coherent state to take into account the dilution of the number density by expansion. Since the 
scalar fields  $\phi_{1}$ and $\phi_{2}$ are independent fields, the coherent state of the complex field is a product of the coherent states for $\phi_{1}$ and $\phi_{2}$, $|\Phi(t)> = |\phi_{1}(t)>|\phi_{2}(t)>$. Therefore, with the baryon number density operator given by $\hat{n} =   \hat{\dot{\phi}}_{1} \hat{\phi}_{2} - \hat{\dot{\phi}}_{2} \hat{\phi}_{1} $, the expectation value of the baryon asymmetry in the coherent state is given by 
\be{cs2} <\Phi(t)|\hat{n} |\Phi(t)>      =   <\Phi(t)|\hat{\dot{\phi}}_{1} \hat{\phi}_{2} - \hat{\dot{\phi}}_{2} \hat{\phi}_{1}|\Phi(t)>  = \dot{\phi}_{1,\;cl}\phi_{2,\;cl} - \dot{\phi}_{1,\;cl}\phi_{1,\;cl} \equiv n_{cl}   ~.\ee 
Therefore the expectation value of the baryon number density operator is equal to the baryon number density $n_{cl}$ calculated using the classical fields. 
When $\phi_{i} < m_{\Phi}$, the variance of the field in the coherent state will become large compared to the squared classical field. Therefore there will be large quantum fluctuations of the fields about their expectation values and so the field cannot be considered classical. However, the correlation length of the quantum fluctuations cannot be larger than the horizon at inflaton decay. Since the volume that evolves into the presently observed Universe will be very much larger than the horizon volume at inflaton decay, the observed baryon asymmetry will be given by its spatial average value and so will equal the expectation value of the baryon asymmetry. Therefore the baryon asymmetry will equal its classical value even when $\phi_{i} < m_{\Phi}$. 
This shows that it is the coherent state describing the scalar 
field, rather than its classical nature, that is essential for AD baryogenesis.  

In reaching this conclusion we have assumed that the mean asymmetry transferred from the condensate by decay is equal to the mean asymmetry in the coherent state of the condensate and that there is no additional washout effect due to the decay process. Condensate decay in this model occurs when the occupation number is less than one, therefore the conventional classical analysis based on production of particles due to a time-dependent classical field is no longer valid. Whilst there is no obvious reason to expect an additional source of washout due to the decay process when the coherent state is no longer in the classical limit, this should be confirmed by a full quantum field theory analysis.  

\section{Conclusions} 

We have presented a new minimal approach to baryogenesis which is based on $B$-violating mass terms for the inflaton. The resulting model requires only the addition of $B$-violating mass terms to an existing inflaton potential and therefore can easily be realised in a wide range of inflation models.  The asymmetry is generated at late times during inflaton oscillations, in contrast to existing inflaton-based AD baryogenesis models which generate the asymmetry during or shortly after inflation. The model also provides exact analytical expressions for the resulting baryon asymmetry.  

In this analysis we have not addressed the question of baryon isocurvature perturbations. We note that these can easily be controlled by including a $\Phi^{4} + \Phi^{\dagger\; 4}$ term in the potential which is significant during inflation and becomes negligible after inflation, whilst leaving open the possibility of observable isocurvature perturbations. A detailed implementation of the mechanism to baryogenesis from AD leptogenesis via inflaton decay, including a discussion of isocurvature perturbations, will be presented in a future work \cite{aj2}. The model also raises new questions regarding the Affleck-Dine mechanism in the limit where the classical approximation is no longer valid, which requires a dedicated analysis.

\section*{Acknowledgements}

The work of ALS is supported by STFC.

\end{document}